\documentclass[11pt]{article}
\usepackage{amssymb}

\newfont{\goth}{eufm10 scaled\magstep1}
\newfont{\tenmsb}{msbm10 scaled\magstep1}
\let\ssection=\section
\renewcommand{\section}{\setcounter{equation}{0}\ssection}

\newcommand{\half}{{\scriptstyle{\frac{1}{2}}}}
\newcommand{\cL}{{\cal L}}

\newcommand{\bR}{{\bf R}}

\newcommand{\vx}{{\vec x}}
\newcommand{\D}{
{D\mkern-2mu\llap{{\raise+0.5pt\hbox{\big/}}}\mkern+2mu}
}

\def\parag{\hfil\break} 
\def\kikezd{\parag\underbar}
\def\p{{\partial}}
\def\vb{{\vec b}}

\def\vbeta{{\vec\beta}}
\def\vgamma{{\vec\gamma}}

\def\vg{{\vec k}}
\def\vj{{\vec\jmath}}
\def\vp{{\vec p}}

\def\vQ{{\vec Q}}
\def\vq{{\vec q}}

\def\vX{{\vec X}}
\def\vj{{\vec\jmath}}

\def\vg{{\vec g}}

\def\vv{{\vec v}}
\def\vx{{\vec x}}

\def\vu{{\vec u}}
\def\vnabla{{\vec\nabla}}

\newcommand{\fg}{\mathfrak{g}}

\begin{document}

\setlength{\baselineskip}{16pt}

\title{Spin and exotic Galilean symmetry}

\author{
C.~Duval
\\
Centre de Physique Th\'eorique, CNRS\\
Luminy, Case 907\\
F-13288 MARSEILLE Cedex 9 (France)
\\
\\
P.~A.~Horv\'athy
\\
Laboratoire de Math\'ematiques et de Physique Th\'eorique\\
Universit\'e de Tours\\
Parc de Grandmont\\
F-37200 TOURS (France)
\\
}

\date{\today}

\maketitle

\begin{abstract}
   A slightly modified and regularized version of the non-relativistic
   limit of the relativistic anyon model
   considered by Jackiw and Nair yields particles
   associated with the twofold central extension of the
   Galilei group, with independent spin and exotic structure.

\end{abstract}

\vskip3mm
\texttt{hep-th/0209166}


\section{Introduction}

Strange things can indeed happen in the plane: for example,
a particle (called an anyon)
can have fractional spin and intermediate statistics
\cite{anyons}. Another one is that the planar Galilei group
admits a non-trivial two-parameter central
extension \cite{centralex1} leading to an ``exotic'' model
(equivalent to non-commutative mechanics) \cite{DH,NCQM}.

  Soon after its introduction, Jackiw and Nair \cite{JaNa2} rederived
our ``exotic'' model in \cite{DH} by taking the non-relativistic limit
of their  relativistic spinning anyon \cite{JaNa}.
This may suggest that spin and exotic structure are related:
  one can trade one for the other, but one cannot
have both. Here, we argue that, at least  non-relativistically,
spin and ``exotic'' structure can coexist independently.
We show this, first, by reviewing the most general group-theoretical
construction associated with the the twofold extended Galilei group,
next, by slightly modifying the contraction considered by Jackiw and Nair
\cite{JaNa2}. Our abstract construction is
illustrated by the acceleration-dependent model of
\cite{LSZ} and by Moyal field theory \cite{MoyalFT, HoMa}.

\goodbreak

\section{Exotic models}

The free ``exotic'' particle model of \cite{DH} consists of a
five-dimensional ``evolution space'' $T^*\bR^2\times\bR$
described by position, $\vx$, momentum,
$\vp$, and time, $t$, (see \cite{SSD}) which is endowed with the presymplectic
two-form
\begin{equation}
\omega
=
-\,dp_i\wedge dx_i
-
\displaystyle\frac{\kappa}{2m^2}\,\epsilon_{ij}\,dp_i\wedge{}dp_j
+dh\wedge dt,
\qquad
\mathrm{with}
\qquad
h=\displaystyle\frac{\vp{\,}^2}{2m},
\label{oursymplectic}
\end{equation}
where $i,j=1,2$ and $m,\kappa$ are given constants interpreted as mass and
``exotic'' parameter.\footnote{As a rule, in the non-relativistic
theory, all spatial indices $i,j,k$ are lower indices.  Ordinary convention is,
however, adopted for the position of the relativistic
indices $\lambda,\mu,\nu$.}
The equations of motion are then given by the null foliation of $\omega$.

The manifest Galilean invariance of the two-form (\ref{oursymplectic})
provides us, via the (pre)\-symplectic version of N{\oe}ther's theorem
\cite{SSD}, with conserved quantities, most of which take the standard form.
The angular momentum, $j$, and the Galilean boost,
$\vg$, contain, however, new terms, namely
\begin{equation}
\begin{array}{ll}
j=\vq\times\vp+\displaystyle\frac{\kappa}{2m^2}\,\vp{\,}^2,
\\[8pt]
g_i=mx_i-p_it+\displaystyle\frac{\kappa}{m}\,\epsilon_{ij}p_j.
\end{array}
\label{ourconserved}
\end{equation}
Those terms change the Poisson brackets of the Galilean boosts, which now
satisfy
$
\{g_i, g_j\}=-\kappa\,\epsilon_{ij}
$
rather than commute, as usual.

The system is {\it bosonic}:
the quantum angular momentum operator is,  in the momentum representation,
$
\widehat{\jmath} =-i\epsilon_{jk}p_{j}\p_{p_{k}}
$
\cite{centralex1,DH,GRIG}.
Owing to the inherent ambiguity
of planar angular momentum, an arbitrary constant, $s_{0}$,
representing anyonic spin, can be freely added
to the angular momentum, though.
Let us therefore discuss all related classical models.


All ``elementary'' systems  associated with the doubly-centrally
extended Galilei group were
determined by Grigore \cite{GRIG} who,  following Souriau
\cite{SSD}, identifies them with
 coadjoint orbits of the group, endowed with their
canonical symplectic structure. (Here the word ``elementary''
means that the action of the Galilei group should be transitive;
at the quantum level this means that the representation be
irreducible.)

These orbits are $4$-dimensional, and depend on four
parameters denoted by $s_{0}$, $h_0$, $m$ and~$\kappa$. Their
symplectic structure are pulled-back, on ``evolution space'', precisely as the
closed two-form (\ref{oursymplectic}). This latter only depends on
$m$ and $\kappa$; the two other parameters only show up
in the way the Galilei group acts on the orbit, i.e.,
in the associated conserved quantities. These latter are
found to be the linear momentum, $\vp$, the boost, $\vg$, in
(\ref{ourconserved}), together with the angular momentum and energy, viz
  \goodbreak
\begin{equation}
\begin{array}{c}
j=\vx\times\vp+\displaystyle\frac{\kappa}{2m^2}\vp{\,}^2+
s_{0},
\\[8pt]
h=\displaystyle\frac{\vp{\,}^2}{2m}+h_0,\hfill
\end{array}
\label{EXconsquant}
\end{equation}
supplemented with the Casimir invariants $m$ and $\kappa$.
The parameter $s_{0}$ is, hence, interpreted as {\it anyonic spin}, and
$h_0$ is {\it internal energy}.  Let us insist that $s_0$ and the exotic
parameter $\kappa$ are, {\it a priori}, independent quantities.
Plainly, the model of \cite{DH}
realizes (\ref{EXconsquant}) with $s_0=h_0=0$.

 Nonvanishing spin and internal energy also arise for the
acceleration-dependent
system considered by Lukierski et al. \cite{LSZ}. Here, phase space is
$6$-dimensional with coordinates~$\vX, \vp,\vQ$, and
endowed with the following symplectic structure and Hamiltonian
\begin{equation}
\begin{array}{ll}
\Omega=-dp_i\wedge dX_i
-
\displaystyle\frac{\theta}{2}\epsilon_{ij}dp_i\wedge dp_j
-
\displaystyle\frac{1}{2\theta}\epsilon_{ij}\,dQ_i\wedge{}dQ_j,
\\[6pt]
h=\displaystyle\frac{\vp{\,}^2}{2m}-\displaystyle\frac{\vQ^2}{2m
\theta^2},
\end{array}
\label{LSZsympham}
\end{equation}
respectively. This system is also invariant with respect to the planar Galilei
group; the associated conserved quantities, namely the linear momentum, $\vp$,
the boost, $\vec{g}$,
the energy~$h$ in (\ref{LSZsympham}), augmented with the angular
momentum,
\begin{equation}
\begin{array}{ll}
j=\vX\times\vp+\displaystyle\frac{\theta}{2}\vp{\,}^2
+
\displaystyle\frac{1}{2\theta}\vQ^2,\hfill
\end{array}
\label{LSZconsquant}
\end{equation}
provide us with a  more general moment map
(\ref{EXconsquant})
with internal energy $h_{0}=-\vQ^2/(2m\theta^2)$ and anyonic
spin $s=\vQ^2/(2\theta)$.
 Once again,
the  exotic parameter $\kappa=-m^2\theta$ is non-trivial.
Let us mention that the spin  can
 be made totally independent from the exotic parameter
by adding a suitable term,
whereas the constraint $s=-m\theta h_{0}$
is also relaxed. See the third reference in \cite{LSZ}.
\goodbreak

It is worth pointing out that the appearance of these two models is
consistent with Souriau's  general  ``d\'ecomposition barycentrique''
(Theorem (13.15) in   \cite{SSD}). Let us indeed consider
the planar Galilei group $G$.
The translations and the boosts span an invariant abelian subgroup
$\widetilde{G}$ of  $G$ (with Lie algebra $\widetilde{\fg}\subset\fg$), so that
the quotient group, $G_{0}=G/\widetilde{G}$, is the direct product of rotations
and time translations.

Let us consider a classical system represented by
a manifold $M$ endowed with a symplectic two-form,
$\Omega$,  upon which the (planar) Galilei
group $G$ acts by symmetries, i.e., in a Hamiltonian fashion. Then
the direct product $G\times G_{0}$ is also a symmetry \cite{SSD}.
Now, if dim $M = 4$, and the symplectic action is transitive,
then $M$ can be identified with an affine-coadjoint orbit endowed with
its canonical symplectic structure.
In fact, $M$ is the dual of the invariant abelian Lie
subalgebra~$\widetilde\fg$,
viz $M\cong\widetilde\fg^*$.
In our case, we get precisely
the free exotic particle model (\ref{oursymplectic}), considered in \cite{DH}.
If, however, dim$M\geq6$, then
the system decomposes into the direct product of
our ``free exotic particle'' phase space and a symplectic manifold,
$M_{0}$, which describes the internal motions
(this latter is distinguished by
the vanishing of the ``standard'' conserved quantities), namely
\begin{equation}
 M\cong\widetilde\fg^*\times M_{0}.
\label{intdecomp}
\end{equation}
The extended group
$G\times G_{0}$ respects this decomposition: the rotations and time
translations act independently on the internal space,
contributing (anyonic) spin and internal energy.

This is exactly what happens for the extended model of
\cite{LSZ}, whose phase space is decomposed into that of our
``free exotic particle'' times the symplectic plane. (An additional
constraint links the rotations with the time translations, as highlighted
by the linked extra terms in (\ref{LSZsympham}) and (\ref{LSZconsquant}).)

A field theoretical model with anyonic spin and non-trivial exotic
structure is provided by
the Moyal-Schr\"odinger field theory \cite{MoyalFT} given by the
non-local Lagrange density
\begin{equation}
L_{NC}=
\frac{i}{2}\left(\bar{\psi}\star\p_t\psi-
\p_t\bar{\psi}\star\psi\right)
-
\frac{1}{2m}\vnabla\bar{\psi}\star\vnabla\psi
\label{freeNClag}
\end{equation}
with the Moyal star-product associated with the deformation
parameter $\theta$. As shown recently~\cite{HoMa},
the theory is Galilei invariant, and the boosts
\begin{equation}
     {\cal G}_{i}=m\int\!x_{i}(\bar{\psi}\star\psi)\, d^{2}\!\vx
     -t{\cal P}_{i}
     -\frac{1}{2} m\theta\epsilon_{ij}{\cal P}_{j},
     \label{xboost}
\end{equation}
(where
$
\vec{\displaystyle{\cal P}}=\displaystyle{\int}\vj\,d^{2}\!\vx
$, with
$
\vj=
\big[\bar{\psi}(\vnabla\psi)-(\vnabla\bar{\psi})\psi\big]/(2i)
$
is the conserved momentum) satisfy the ``exotic'' commutation relation
$\{{\cal G}_1,{\cal G}_2\}=-\kappa\int\vert\psi\vert^2d^{2}\!\vx$ with
$\kappa=-m^2\theta$.
Note that the angular momentum
  \begin{equation}
     {\cal J}=
     \int\left[\vx\times\vec{\jmath}
     -\frac{\theta}{2}\vert\vnabla{\psi}\vert^2
     +s_0\vert\psi\vert^2
     \right] d^{2}\!\vx
     \label{xangmom}
\end{equation}
is anyonic, cf. (\ref{EXconsquant}).

Yet another illustration is provided by the first-order
non-local model given by the Lagrangian
\begin{equation}
   \Im\left\{\psi^\dagger\star\Big(\half(1+\sigma_{3})\p_{t}
    -\vec{\sigma}\cdot\vnabla-im(1-\sigma_{3})\Big)\psi\right\}
    \label{MLL}
\end{equation}
where $\psi=\left(\begin{array}{c}\Phi\\ \chi
\end{array}\right)$
is a two-component Pauli spinor.
Despite the presence of the Moyal product, the associated
Euler-Lagrange equation is the
``non-relativistic Dirac equation''  of L\'evy-Leblond \cite{LL},
\begin{equation}
    \begin{array}{ll}
    (\p_{1}+i\p_{2})\Phi+2im\chi&=0
    \\[8pt]
    \p_{t}\Phi-(\p_{1}-i\p_{2})\chi&=0.
    \end{array}
    \label{LLeq}
\end{equation}
These are Galilei invariant when the boosts are implemented as \cite{LL}
\begin{equation}
    \psi(\vx,t)=\left(\begin{array}{cc}1&0\\
    -\half (b_{1}+ib_{2})&0
    \end{array}\right)
	\exp im[\vb\cdot\vx-t\vb{\,}^2/2]\psi(\vx-\vb t,t).
    \label{boostLL}
    \end{equation}
Using the same technique as in \cite{HoMa}, we find that the  Moyal
star results in a new term in the associated conserved quantity, viz.
\begin{equation}
     {\cal G}_{i}=m\int\!x_{i}(\bar{\Phi}\star\Phi)\, d^{2}\!\vx
     -t{\cal P}_{i}
     -\frac{1}{2} m\theta\epsilon_{ij}{\cal P}_{j},
     \label{LLboost}
\end{equation}
where
$
\vec{\displaystyle{\cal P}}=
\displaystyle{\int}\big[\bar{\Phi}(\vnabla\Phi)-
(\vnabla\bar{\Phi})\Phi\big]/(2i)\,d^{2}\!\vx
$,
cf. (\ref{xboost}). The ${\cal G}_{i}$'s satisfy the ``exotic''
commutation relation.
\goodbreak

\section{Non-relativistic limit by group contraction}\label{NRlimit}

Let us now turn to the non-relativistic limit
of the relativistic anyon model \cite{JaNa} considered by
Jackiw and Nair in \cite{JaNa2}.
Expanding  to first order in $1/c$ the restriction to
$p_0\simeq mc^2+\vp{\,}^2/(2m)$,
the relativistic symplectic structure on $6$-di\-men\-sio\-nal phase space
\begin{equation}
\Omega_{\mathrm{JN}}=-dp_\mu\wedge dx^\mu+\frac{Sc^2}{2}
\epsilon^{\mu\nu\rho}\,\frac{p_\mu{}dp_\nu\wedge{}dp_\rho}
{(p^2)^{3/2}}
\label{JNsymp}
\end{equation}
(where the Greek indices range from
$0$ to $2$) yields indeed our
presymplectic two-form (\ref{oursymplectic}), when their ``spin'', $S$, is
identified with our coefficient $\kappa$.
\goodbreak

The relativistic model admits, furthermore, the symmetry with generators
\begin{equation}
J_\mu=
\epsilon_{\mu\nu\rho}x^{\nu}p^\rho
+Sc^2\frac{p_{\mu}}{\sqrt{p^2}}
\end{equation}
which satisfy the o$(1,2)$ commutation relations relations,
$\{J^\mu,J^\nu\}=\epsilon^{\mu\nu\rho}J_\rho$.
Then the quantities
$
\widetilde{g}_i=\epsilon_{ij}J_j/c,
$
satisfy, to leading order in $1/c$,
the exotic commutation relation above,
allowing Jackiw and Nair to identify their $\widetilde{g}_i$
with the Galilean boosts.
In particular, $S$ becomes the exotic parameter
$\kappa$.
Their angular momentum,
\begin{equation}
J_0\simeq\epsilon_{ij}x_ip_j+Sc^2+\frac{S}{m^2}\,\vp{\,}^2,
\label{JNangmom}
\end{equation}
does not admit, however, a well-defined limit  as $c\to+\infty$, unless the
divergent constant  $Sc^2$ is removed. Only at this price are the exotic
quantities, namely the angular momentum and the boosts, (\ref{ourconserved}),
recovered.


Let us now present a mathematical construction that allows us to absorb the
divergent terms into some extra coordinates. Extending the approach of
Jackiw and
Nair \cite{JaNa2}, our clue is to work with differential forms and group
actions,
rather than with the associated conserved quantities.

To get an insight, let us first discuss the model of a
massive, spinless, relativistic particle in the plane.
Following Souriau \cite{SSD}, it  corresponds to a certain coadjoint orbit
of the
Poincar\'e group (we still denote by $G$), symplectomorphic to
$T^*\bR^2$. The pull-back of its canonical symplectic structure, $\omega$,
to~$G$ is indeed $d\alpha$, where
$\alpha=p_{\mu}dx^{\mu}$ is a one-form on $G$.
Using physical coordinates in a Lorentz frame, we obtain
\begin{equation}
\alpha=-\vp\cdot d\vx+mc^2\sqrt{1+\frac{\vp{\,}^2}{m^2c^2}}\,dt,
\label{alphaPoincareBis}
\end{equation}
which is the ``Cartan-form'' \cite{SSD} of the free relativistic Lagrangian,
$\cL_{0}=mc^2\sqrt{1-\vv{\,}^2/c^2}$.

\goodbreak

Let us emphasize that the one-form $\alpha$
(as well as the Lagrangian $\cL_{0}$) diverges
in the Galilean limit $c\to\infty$, as clearly seen by writing
(\ref{alphaPoincareBis}) as
\begin{equation}
\alpha
=
-\vp\cdot d\vx
+\left[
mc^2
+
\frac{\vp{\,}^2}{2m}+\mathcal{O}\left(\frac{1}{c^2}\right)
\right]dt.
\label{alphaPoincareTer}
\end{equation}
The two-form $d\alpha$ has,
nevertheless, a well-behaved limit, namely the familiar presymplectic two-form
of an ``ordinary'' free, spinless, non-relativistic particle of mass
$m$, namely
\begin{equation}
d\alpha
\approx
-dp_i\wedge{}dx_i
+
dh\wedge{}dt
\qquad
\mathrm{with}
\qquad
h=\frac{\vp{\,}^2}{2m},
\label{limitdalphaPoincare}
\end{equation}
where the notation ``$\approx$'' stands for ``up to higher-order terms in
$1/c^2$''.

In order to cure the pathology of (\ref{alphaPoincareTer}), we
consider the  {\it trivial} central extension
$\widehat{G}=G\times\bR$ of the planar Poincar\'e
group $G$. The new one-form to consider on~$\widehat{G}$ is the
left-invariant one-form
\begin{equation}
\widehat{\alpha}=\alpha+\varepsilon d\tau
\label{alphahat}
\end{equation}
where $\tau$ is a coordinate on the centre  $(\bR,+)$ and
$\varepsilon\in\bR$.
We now posit, for the sake of convenience,
$
\tau=t+{u}/{c^2}
$
where $u$ is a new real parameter that replaces our old
$\tau$. Also, let us introduce a new real constant, $h_0$,
via
\begin{equation}
\varepsilon=-mc^2+h_0.
\label{intenergydec}
\end{equation}
Then the divergence in the one-form
$\widehat{\alpha}$ disappears (cf. (\ref{alphaPoincareTer})),
and the latter simply reads
\begin{equation}
\widehat{\alpha}
\approx
-\vp\cdot d\vx
+\left[\frac{\vp{\,}^2}{2m}
+h_0\right]dt
-
m\,du.
\label{alphaPoincareQuarto}
\end{equation}

\goodbreak
The one-form $\widehat{\alpha}$ hence converges in the limit
$c\to+\infty$. Note that we also get the
``internal energy'', $h_0$, specific to Galilean Hamiltonian mechanics.
It is worth mentioning that (\ref{intenergydec}) is the most general
Ansatz for a power series in~$c^2$ that guarantees convergence
of~$\widehat{\alpha}$ while bringing non-trivial contribution to the
non-relativistic limit.

Thus, the divergent term in $\alpha$ is absorbed into
an exact one-form that involves the extra coordinate that drops out.
The remaining part is regular and yields the correct conserved quantities.


The advantage of this approach is that trivially extended
Poincar\'e group now contracts nicely
to the singly extended Galilei group.
The infinitesimal action of the
Poincar\'e group $G$ on space-time reads, in fact,
\begin{equation}
\left\{
\begin{array}{ccc}
\delta x_{i}\hfill&=&\omega{}\epsilon_{ij}x_{j}+\beta_{i}{}t+\gamma_{i},\hfill
\\[8pt]
\delta{t}\hfill&=&\displaystyle\vbeta\cdot\vx/{c^2}+\epsilon,\hfill
\end{array}
\right.
\label{poincareAction}
\end{equation}
with
$\omega\in\bR$ generating rotations,
$\vbeta\in\bR^2$ boosts,
$\vgamma\in\bR^2$ space translations and $\epsilon\in\bR$ time
translations.

\goodbreak

The infinitesimal action of the centre $(\bR,+)$ of the extended group
$\widehat{G}$ is readily written as
$
\delta{\tau}={\eta}/{c^2}+\epsilon'
$
with $\eta,\epsilon'\in\bR$, so that the above definition of $u$ yields, in the
limit $c\to+\infty$,
$\epsilon'=\epsilon$ and
\begin{equation}
\delta{u}=-\vbeta\cdot\vx+\eta,
\label{BargmannAction}
\end{equation}
corresponding precisely to the infinitesimal action of the non-trivial
$1$-parameter central extension of the Galilei group
$\widehat{G}_\infty$ on the ``vertical'' fiber parametrized by $u$, while the
usual Galilei action
on spacetime is plainly deduced from (\ref{poincareAction}) in the limit
$c\to+\infty$, viz
\begin{equation}
\left\{
\begin{array}{cc}
   \delta x^i\hfill=&
\omega{}\epsilon_{ij}x_{j}+\beta_{i}{}t+\gamma_{i},
\hfill\\[8pt]
\delta{t}\hfill=&
\epsilon.
\hfill
\end{array}
\right.
\label{galileiAction}
\end{equation}

\section{Spin and exotic parameter}

Let us now extend our improved contraction to spin.
Starting with spinning massive particles
moving in Minkowski spacetime $\bR^{2,1}$, a similar procedure leads
to three items analogous to those previously highlighted.


  For a particle with
spin $s\in\bR$ and mass $m>0$, dwelling in Minkowski spacetime,
the one-form to start with on the Poincar\'e group, $G$, is given \cite{SSD} by
$
\alpha=p_\mu{}dx^\mu+s\,e_{1\mu}\,de_2^\mu
$
where $p^\mu=m{}e_0^\mu$ and $(e_0,e_1,e_2)_x$ is an orthonormal Lorentz
frame at the point $x\in\bR^{1,2}$. This one-form~
retains, in an adapted coordinate system, the rather complicated-looking
expression
\begin{equation}
\begin{array}{rcl}
\alpha
&=&
\displaystyle
-\vp\cdot{}d\vx+mc^2\sqrt{1+\frac{\vp{\,}^2}{m^2c^2}}\,dt
\\[12pt]
&+&s\,
\displaystyle
\frac{\sqrt{1+\displaystyle\frac{\vp{\,}^2}{m^2c^2}}}
{1+\displaystyle\frac{(\vp\times\vu)^2}{m^2c^2}}
\Bigg[
\left(1+\displaystyle\frac{\vp{\,}^2}{m^2c^2}\right)\vu\times{}d\vu
+
\displaystyle\frac{\vp\cdot\vu}{m^2c^2}\,d\vu\times\vp
\\[20pt]
&&
+\displaystyle\frac{\vu\times\vp\,(\vp\cdot\vu)\,\vp\cdot{}d\vp}
{m^4c^4\left(1+\displaystyle\frac{\vp{\,}^2}{m^2c^2}\right)}
-
\displaystyle\frac{\vu\times\vp}{m^2c^2}\,d(\vp\cdot\vu)
\Bigg]
\label{alphaSpinPoincare}
\end{array}
\end{equation}
with $\vu\in{}S^1$, an arbitrary unit vector in the plane.
Note that $\vu\times{}d\vu=d\phi$, where $\phi$ is
the argument of $\vu$ (actually
   the rotation angle of the $\mathrm{SO}(2)$ subgroup of the Lorentz
group $\mathrm{SO}(1,2)$).
A tedious calculation yields furthermore a presymplectic two-form,
$\omega=d\alpha$, similar to (\ref{JNsymp}),
whose behaviour to order $c^{-2}$ is
\begin{equation}
\omega
\approx
-dp_i\wedge{}dx_i
+
dh\wedge{}dt
+
\frac{s}{m^2c^2}\,dp_1\wedge{}dp_2
\label{sigmaBis}
\end{equation}
where $h$ is as in (\ref{oursymplectic}).


Note that if one
considers $s$ as being independent of $c$, then the two-form
$\omega$ in (\ref{sigmaBis}) plainly tends, as $c\to+\infty$, to that
of an ``ordinary'' non-relativistic particle, (\ref{limitdalphaPoincare}).
Our clue is to posit, instead, the Jackiw-Nair-inspired Ansatz, cf.
(\ref{intenergydec}),
\begin{equation}
s=
\kappa{}c^2+s_0
\label{theAnsatz}
\end{equation}
where  $\kappa$ and $s_0$ are new constants.
 Then, in the non-relativistic limit  we recover precisely our exotic two-form
(\ref{oursymplectic}). This is just like in
\cite{JaNa2}, up to a minor difference in the interpretation: it is our
$s$, and
not their $S=sc^2$ in (\ref{JNsymp})
that should be called relativistic spin.
The parameter~$S$  has indeed physical dimension
$[S]=[\hbar/c^2]$,
and cannot represent spin, whose dimension, $[\hbar]$, is carried
correctly by our $s$. The constants $\kappa$ in (\ref{theAnsatz})
(whose dimension is $[\hbar/c^2]$)
is hence interpreted as the exotic parameter; also
$s_0$ will turn out to be Galilean anyonic spin.

Presenting the spin term in equation (\ref{alphaSpinPoincare}) as
\begin{eqnarray}\label{limitofIdJ}
\kappa{}c^2d\phi
+
s_0\,d\phi
+
\frac{\kappa}{m^2}
\Bigg[
\left(
\frac{3}{2}\,\vp{\,}^2 - (\vp\times\vu)^2
\right)d\phi
+(\vp\cdot\vu)\,\,d\vu\times\vp
-
\vu\times\vp\,d(\vp\cdot\vu)
\Bigg]
+
\mathcal{O}\left(\frac{1}{c^2}\right)
\end{eqnarray}
yields
\begin{equation}
\alpha
\approx
-\vp\cdot d\vx
+\left[mc^2+\frac{\vp{\,}^2}{2m}\right]dt
+\kappa{}c^2d\phi
+s_0d\phi
+
\frac{\kappa}{2m^2}\,\vp\times{}d\vp
\label{alphaLimit}
\end{equation}
modulo a closed one-form. Thus, while
   the exterior derivative $d\alpha$  behaves correctly,
the one-form (\ref{alphaLimit}) contains instead
two divergent terms, namely $mc^2dt$ and
$\kappa{}c^2d\phi$. Removing them and setting $s_{0}=0$
would yield the Cartan-form of the Lagrangian used in \cite{DH}.

This time, regularization
is achieved by the (trivial) {\it double central extension}\
$\widehat{\!\widehat{G}}=\widehat{G}\times\bR$
of the Poincar\'e group, endowed with the canonical one-form:
$
\widehat{\!\widehat{\alpha}}
=
\widehat{\alpha}
+
\chi{}d\theta
$
where $\widehat{\alpha}$ is as in~(\ref{alphahat}) and $\theta$ parametrizes
the new $(\bR,+)$-factor, the coordinate $\chi$ having the dimension
of~$\hbar$.
  The divergence associated with the energy having been removed by the first
trivial central extension, let us posit\footnote{Adding a
$c$-independent constant to
$\chi$ would just modify the form of the (anyonic) spin $s_0$ by an overall
additive constant.}
$
\chi=-\kappa{}c^2
$
and
$
\theta=\phi-{w}/{c^2}
$
where $w\in\bR$ is the new parameter to consider in place of $\theta$.
Then, unlike $\alpha$ in (\ref{alphaLimit}), the one-form
$\widehat{\!\widehat{\alpha}}$
{\it is} well behaved
in the limit $c\to+\infty$ and retains, (modulo a closed one-form),
the expression
\begin{equation}
\widehat{\!\widehat{\alpha}}
\approx
-\vp\cdot d\vx\,
+\left[
\frac{p^2}{2m}+h_0
\right]dt
-m\,du
+s_0d\phi
+
\frac{\kappa}{2m^2}\,\vp\times{}d\vp
+\kappa{}dw.
\label{alphaLimitOK}
\end{equation}


In order to make sense of the group contraction
$
\widehat{\!\widehat{G}}_\infty=\lim_{c\to+\infty}\widehat{\!\widehat{G}},
$
we need to compute the infinitesimal action of the Lorentz group on the
rotations, parametrized by~$\phi$, which serve to define our second extension
coordinate, $w$.
Using the known form of the matrices spanning
the Lie algebra $\mathrm{o}(1,2)$, one writes the infinitesimal
Lorentz action
on itself to obtain  finally
\begin{equation}
\delta\phi
\approx
\omega+\frac{1}{2c^2}\,\vbeta\times\vb.
\label{deltaphi}
\end{equation}
The infinitesimal action of the extra central subgroup $(\bR,+)$ of
$\,\widehat{\!\widehat{G}}$
being written as
$
\delta{\theta}={\varrho}/{c^2}+\omega'
$
with $\varrho,\omega'\in\bR$, we finally get from the definition of
$w$, in the limit
$c\to+\infty$, firstly $\omega'=\omega$ and, secondly, the
expression
\begin{equation}
\delta{w}=\varrho+\frac{1}{2}\,\vbeta\times\vv
\label{exoticExtensionAction}
\end{equation}
where we have put $\vv=\vp/m$ for the
velocity.

Formula (\ref{exoticExtensionAction}) gives the
infinitesimal action of the exotic Galilei
group $\,\widehat{\!\widehat{G}}_\infty$ on the new, exotic,
extension fiber with coordinate $w$. This latter can be viewed as providing an
extended evolution space.
Notice, though, that the action of
$\,\widehat{\!\widehat{G}}_\infty$ on this evolution space---whose
infinitesimal
form is given by (\ref{galileiAction}),
(\ref{BargmannAction}), and
(\ref{exoticExtensionAction})---is {\it not} the lift of an
action on some extended space-time, as it involves the velocity~$\vv$ in
(\ref{exoticExtensionAction}).

Let us mention that the conserved quantity associated to a symmetry
is obtained, in this framework, by simply contracting the (invariant)
one-form $\widehat{\!\widehat{\alpha}}$ on the group with the
infinitesimal generator of the
 symmetry. Using (\ref{alphaLimitOK}),
we recover in particular the general conserved
quantities in (\ref{EXconsquant}). No divergences occur.

Our modified contraction  with the  Ansatz (\ref{theAnsatz})
yields hence a non-relativistic model
with both an exotic structure and anyonic spin.

\section{Discussion}

The derivation of the exotic model presented by Jackiw and Nair
\cite{JaNa2} has
some subtle points: their ``spin'', which becomes our exotic parameter, has
not the expected physical dimension, and their angular momentum diverges as
$c$ tends to infinity. Such divergences are indeed familiar in the theory
of group contraction; our mathematical construction, here, allows us to
eliminate
the divergent terms by resorting to a double group extension of the Poincar\'e
group and to recover, for free, the twice centrally extended Galilei group of
planar physics.

In a recent paper, Hagen \cite{Hagen} also discusses the relation
of spin and exotic extension. He considers the
L\'evy-Leblond equation (\ref{LLeq})
but with ordinary Lagrangian, i.e. (\ref{MLL}) with
ordinary product, which has no  non-trivial second extension \cite{LL}.

In conclusion, let us stress that fractional
spin arises due to the commutativity of the planar rotation
group alone, independently of any further symmetry. This is why one
can have relativistic as well as non-relativistic anyons.
Although the commutative structure of plane rotations plays a r\^ole
for the ``exotic'' structure also, this latter clearly involves more
symmetries, namely the two-dimensional group-cohomology of the Galilei group.
On the other hand, this phenomenon definitely does not arise in the
relativistic context since the Poincar\'e group has trivial group-cohomology.

\goodbreak


\kikezd{Acknowledgement:}
We are indebted to Professors R.~Jackiw
and M.~Plyushchay
for their interest and correspondence.




\end{document}